\documentclass[manuscript]{aastex631}

\usepackage{longtable,tabularx,hyperref}
\usepackage{booktabs}
\usepackage{float}
\usepackage{amsmath}
\begin{document}

\title[EPIC 248846335]{Asteroseismological analysis of the non-Blazhko RRab star EPIC~248846335 in LAMOST - Kepler$/$ K2 project}

\correspondingauthor{Jian-Ning, Fu}
\email{jnfu@bnu.edu.cn}

\author{Peng Zong}
\affiliation{Institute for Frontiers in Astronomy and Astrophysics, Beijing Normal University, Beijing 102206, People’s Republic of China}
\affiliation{Department of Astronomy, Beijing Normal University, Beijing 100875, People’s Republic of China}

\author{Jian-Ning Fu}
\affiliation{Institute for Frontiers in Astronomy and Astrophysics, Beijing Normal University, Beijing 102206, People’s Republic of China}
\affiliation{Department of Astronomy, Beijing Normal University, Beijing 100875, People’s Republic of China}

\author{Jie Su}
\affiliation{Yunnan Observatories, Chinese Academy of Sciences, Kunming 650216, People’s Republic of China}
\affiliation{Key Laboratory for the Structure and Evolution of Celestial Objects, Chinese Academy of Sciences, Kunming 650216, People’s Republic of China}
\affiliation{International Centre of Supernovae, Yunnan Key Laboratory, Kunming 650216, People’s Republic of China}

\author{Xueying Hu}
\affiliation{Department of Astronomy, Beijing Normal University, Beijing 100875, People’s Republic of China}

\author{Bo Zhang}
\affiliation{National Astronomical Observatories of China, Chinese Academy of Science, Beijing 100012, People’s Republic of China}

\author{Jiaxin Wang}
\affiliation{College of Science, Chongqing University of Posts and Telecommunications, Chongqing 400065, People’s Republic of China}

\author{Gao-Chao Liu}
\affiliation{Center for Astronomy and Space Sciences, China Three Gorges University, Yichang 443002, People’s Republic of China}

\author{Gang Meng}
\affiliation{Department of Astronomy, Beijing Normal University, Beijing 100875, People’s Republic of China}

\author{Gianni Catanzaro}
\affiliation{INAF - Osservatorio Astrofisico di Catania, Via S.Sofia 78, I-95123, Catania, Italy}

\author{Antonio Frasca}
\affiliation{INAF - Osservatorio Astrofisico di Catania, Via S.Sofia 78, I-95123, Catania, Italy}

\author{Haotian Wang}
\affiliation{Department of Astronomy, Beijing Normal University, Beijing 100875, People’s Republic of China}

\author{Weikai Zong}
\affiliation{Department of Astronomy, Beijing Normal University, Beijing 100875, People’s Republic of China}




\begin{abstract}
We conduct an asteroseismological analysis on the non-Blazhko ab-type RR Lyrae star EPIC 248846335 employing the Radial Stellar Pulsations (RSP) module of the Modules for Experiments in Stellar Astrophysics (MESA) based on the set of stellar parameters. The atmospheric parameters as $T_\mathrm{eff}$ = 6933$\pm$70 $K$, log $g$ = 3.35$\pm$ 0.50 and [Fe/H] = -1.18 $\pm$ 0.14 are estimated from the Low-Resolution Spectra of LAMOST DR9. The luminosity $L$ = 49.70$_{-1.80}^{+2.99}$ $L_\odot$ and mass M = 0.56 $\pm$ 0.07 $M_\odot$ are calculated, respectively, using the distance provided by Gaia and the metallicity estimated from the Low-Resolution Spectra. The Fourier parameters of the light curves observed by $K2$ and RV curves determined from the Medium-Resolution Spectra of LAMOST DR10 are also calculated in this work. The period of the fundamental mode of the star and the residuals $r$ of the Fourier parameters between the models and observations serve to select optimal model, whose stellar parameters are $T_\mathrm{eff}$ = 6700 $\pm$ 220 K, log $g$ = 2.70, [Fe/H] = -1.20 $\pm$ 0.2, M = 0.59 $\pm$ 0.05 $M_\odot$, and $L$ = 56.0 $\pm$ 4.2 $L_\odot$. The projection factors are constrained as 1.20 $\pm$ 0.02 and 1.59 $\pm$ 0.13 by the blue- and red-arm observed velocities with their corresponding RV curves derived from the best-fit model, respectively. The precise determination of stellar parameters in ab-type RR Lyrae stars is crucial for understanding the physical processes that occur during pulsation and for providing a deeper understanding of its Period-Luminosity relationship.
\end{abstract}



\section{Introduction} 
\label{sec:intro}
The RR Lyrae variables (RRLs), with masses ranging from 0.5 to 0.8$M_\odot$, are large-amplitude pulsations. They are located at the intersection of the horizontal branch (HB) and the instability strip in the Hertzsprung-Russell (H-R) diagram, where they undergo helium core burning \citep{2010aste.book.....A}. These stars pulsate due to the $\kappa$ mechanism, driven by partial ionization of hydrogen and helium. They are radial pulsating variables with typical pulsation periods between 0.2 and 1 day. RRLs display light variations of 0.3 to 1.7 magnitudes in the $V$ band and have effective temperatures ranging from 6100 to 7400K, corresponding to spectral types A2 to F6. They can be categorized into the following types: RRab stars pulsating in the fundamental mode, RRc stars in the first overtone, and RRd stars in both modes \citep{2021ApJ...922...20B}. The shorter-period RRc stars, occasionally referred to as RRe, represent the metal-rich extension of the RRc class \citep{1997ApJ...483..811B}. Due to their adherence to a precise period-luminosity-metallicity (PLZ) relation, particularly in near-infrared bands, RRLs serve as critical tools for tracing and measuring distances to ancient stellar populations within the Milky Way and nearby galaxies \citep{2001MNRAS.326.1183B,2004ApJS..154..633C,2015ApJ...807..127M,2021ApJ...909..200B}. Moreover, the Blazhko effect, which is the periodic modulation of the light curves' amplitude and phase in RRLs, remains an interesting unsolved problem in astrophysics since its identification \citep{1907AN....175..325B,1916ApJ....43..217S}.

The stellar models of RRLs have been studied for a long time. A study by \cite{2000ApJ...532L.129B} used full-amplitude, nonlinear, convective hydrodynamical models to investigate the behavior of the RRc variable star U Com. The study confirmed that the theoretical models accurately reflect the observed luminosity changes throughout the pulsation cycle. \cite{2005AJ....129.2257M} applied nonlinear convective pulsation models to 14 Large Magellanic Cloud (LMC) RRLs, comprising of an equal number of RRab and RRc stars \citep{2003MNRAS.344.1097B,2003ApJ...596..299M}. This research evaluated the theoretical models and yielded a new independent distance estimate, which significantly impacted the calibration of the RRL distance scale. \cite{2007A&A...474..557M} successfully matched nonlinear pulsation models to the observed light curves of 4 RRc and 2 RRab stars in the Galactic globular cluster M3. This study demonstrated theoretical consistency with observed light curve morphologies and intrinsic stellar parameters, in line with evolutionary expectations for the given metallicity. 

\cite{2013MNRAS.428.3034S} employed nonlinear hydrodynamic pulsation models to explore the stellar parameters of OGLE-BLG-RRLYR-02793 \citep{2012Natur.484...75P}, using light and radial velocity (RV) curves, although this object is not a RR Lyrae star. The radial pulsations of RRLs offer a means to probe hydrodynamic processes through theoretical models, which can be benchmarked against observed light and RV curves to refine stellar parameters. However, acquiring complete RV curves is challenging, especially for fainter stars, due to the extensive telescope time required, as noted by \cite{2013MNRAS.428.3034S}. The LAMOST-$\sl Kepler$/$\sl K2$ surveys (LKS) have provided a wealth of multi-epoch spectra for numerous $\sl Kepler$/$\sl K2$ targets \citep{2015ApJS..220...19D,2018ApJS..238...30Z,2020ApJS..251...27W,2020RAA....20..167F}, enabling the extraction of atmospheric parameters and RV curves for RRLs within the $Kepler$/$K2$ fields. Based on those resources, \cite{2021MNRAS.506.6117W} conducted asteroseismological analyses on the non-Blazhko ab-type star EZ Cnc (EPIC 212182292) using $\sl K2$ light curves and RV data from LKS Medium-Resolution Spectra (MRS). This analysis was performed with the RSP module of the Modules for Experiments in Stellar Astrophysics (MESA) suite \citep{2011ApJS..192....3P,2013ApJS..208....4P,2015ApJS..220...15P,2018ApJS..234...34P,2019ApJS..243...10P,2023ApJS..265...15J}, which simulates large-amplitude, self-excited pulsations as stars transit the instability strip on the H-R diagram. The study not only determined the stellar parameters for EZ Cnc but also estimated the projection factor ($p$ = 1.22), a critical parameter in the Baade-Wesselink (BW) method for distance estimation \citep{2004A&A...428..131N,2017MNRAS.466.2842K,2017A&A...604A.120N}, which converts observed RV variations into pulsation velocities of the stellar photosphere. Notably, the $p$-factor may vary depending on the spectral lines used for RV measurements \citep{2017A&A...604A.120N}. According to \cite{2020RAA....20...51Z}, the MRS from LAMOST's blue-arm predominantly targets the Mg{\sc i}b triplet, while the red-arm spectra capture the H$\alpha$ line, thereby offering a unique opportunity to assess the $p$-factors using RV curves derived from both spectral regions.

In this work, we conduct an asteroseismological analysis of the non-Blazhko RRab-type star EPIC 248846335 ($\alpha_{2000}$ = 10h:48min:11.650s, $\delta_{2000}$ = +11$^{o}$48$^{'}$44.08$^{''}$, $Kp$ = 14.713 mag) to determine the values of the projection factors $p$ and constrain the stellar parameters. We use the RSP module of MESA based on the light curves observed in the $\sl K2$ field, and the RV curves derived from MRS of LAMOST DR10 with atmospheric parameters determined from the Low-Resolution Surveys (LRS) of LAMOST DR9. In Section 2, we present the data collection and analysis. The numerical modeling and discussions are presented in Sections 3 and 4, respectively. Finally, Section 5 provides the conclusions of this paper.

\section{Data Collection and Analysis}
\subsection{Photometry}
\begin{figure*}
\centering
\includegraphics[width=6.2 in]{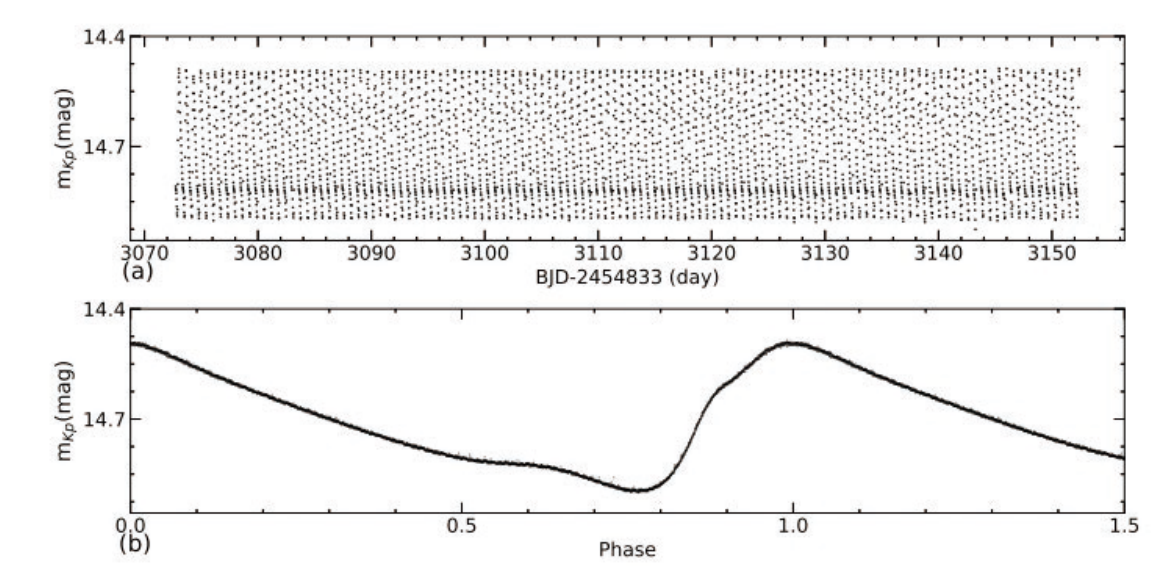}
\caption{(a) Light curves of EPIC 248846335 extracted with LightKurve \citep{2018zndo...1412743B,2021zndo...1181928B}, (b) the phase-folded light curve in the fundamental period.}
\label{fig:LC}
\end{figure*}

The target pixel file (TPF) of EPIC 248846335 obtained with a long cadence observation of $K2$ is downloaded from the Mikulski Archive for Space Telescopes (MAST). All the {\it K2} data used in this paper can be found in MAST \textbf{\citep{2016ApJS..224....2H}}. The package of LightKurve is used to extract light curves from the TPF. To optimize the photometry of the star, several apertures with different pixel sizes are applied to the TPF. After extracting the photometry, the flux is converted to magnitude, and then the light curve is detrended by applying a third-order polynomial and adjusted to the $K{\rm p}$ mean magnitude level given in MAST. The light curve of the star can be seen in Figure~\ref{fig:LC}.

\subsection{Fourier Analysis}
Fourier analysis serves as a potent instrument for exploring the pulsation characteristics of variable stars. The software package Period04 \citep{2005CoAst.146...53L} is employed to perform multifrequency analysis, which applies Fourier transformation combined with least-squares fitting to the light curve, deducing the pulsation frequencies of the star. The main frequency is determined to be $f_0 = 1.5627 (10) \text{ day}^{-1}$, equating to a fundamental period $P_0 = 0.6399 (7) \text{ days}$. Fourier decomposition, first introduced by \cite{1981ApJ...248..291S} to analyze Cepheid light curves, effectively characterizes the light curve features of variable stars. This approach has since become prevalent in the investigation of RRLs \citep{1982ApJ...261..586S, 1993ApJ...410..526S, 2011MNRAS.417.1022N, 2021ApJ...912..144M}. The following Fourier sine series fit the light curve of the target star:
\begin{equation}
\centering
m(t) = A_{0} + \sum_{i=1}^n A_{i} \sin[2\pi if_{0}(t - t_{0}) + \phi_{i}],
\label{eq:1}
\end{equation}
where $m(t)$ denotes the apparent $Kp$ magnitude from $K2$ data, $n$ is the number of harmonic terms, $A_{0}$ the mean $Kp$ magnitude, and $f_{0}$ the fundamental frequency. The variable $t$ corresponds to the time of $K2$ observations (BJD-2454833), with $t_0$ as the epoch of the first maximum. The coefficients $A_i$ and $\phi_i$ represent the amplitude and phase of the $i$-th harmonic, respectively. Following \cite{1981ApJ...248..291S}, certain Fourier coefficients correlate directly with specific physical properties of pulsating stars, typically expressed as linear combinations or ratios of phases and amplitudes:
\begin{equation}
\centering
\phi_{i1} = \phi_{i} - i\phi_{1},
\label{eq:2}
\end{equation}
\begin{equation}
\centering
R_{i1} = \frac{A_{i}}{A_{1}},
\label{eq:3}
\end{equation}
where $i = 2$ or $3$ for the fundamental mode of RRLs \citep{1981ApJ...248..291S}. Corrections for $\phi_{21}$ and $\phi_{31}$ may include integer multiples of $2\pi$ when necessary. The determined pulsation parameters with their corresponding uncertainties are cataloged in the second column of Table~\ref{tab:my_label}. The standard deviation of the residuals of the Fourier decomposition applies to the light curve observed by $K2$ is $\sigma_{LC}$ = 0.008 mag. Adopting the methodology \cite{2023ApJ...945...18Z}, we calculate the total amplitudes $A_{tot}$ and the rise times (RT) of the light and radial velocity curves, with the fitted parameters listed in Table~\ref{tab:my_label}.

\begin{table*}
    \caption{Fourier decomposition parameters of the light curves and radial velocity curves of  EPIC 248846335. ID of those parameters (Column 1), Fourier decomposition parameters of the light curves (Column 2), and Fourier decomposition parameters of radial velocity curves derived from the blue- and red-arm MRS of LAMOST (Columns 3-4), respectively, are presented.}
    \renewcommand\tabcolsep{10pt}
    \begin{tabular}{clll}
    \hline
    \hline
      parameter    & $\sl K2$LC &  RVC (blue-arm) &  RVC (red-arm) \\
         (1)       &    (2)     &  (3) & (4)     \\
    \hline
       A$_1$ &0.160 ($\pm$0.0014)   mag  &   15.75 ($\pm$0.53) Km s$^{-1}$ &  30.69 ($\pm$0.58) Km s$^{-1}$ \\
       A$_{tot}$ & 0.397 ($\pm$0.030)  mag &  43.53 ($\pm$2.01) Km s$^{-1}$ & 66.10 ($\pm$5.26) Km s$^{-1}$\\
       R$_{21}$ & 0.408 ($\pm$0.001)   & 0.32 ($\pm$0.06) & 0.399 ($\pm$0.019)  \\
       R$_{31}$ &  0.219  ($\pm$0.007)   & 0.21  ($\pm$0.08) & 0.218 ($\pm$0.018) \\
       $\phi_{21}$ &2.692 ($\pm$0.023) rad  & 2.35 ($\pm$0.90)  rad  & 2.04 ($\pm$0.57) rad  \\
       $\phi_{31}$ & 5.723 ($\pm$0.017) rad & 5.42 ($\pm$0.81)  rad & 4.17 ($\pm$1.15) rad\\
        RT  &  0.231 ($\pm$0.010) rad &  0.294 ($\pm$0.008) rad  &0.310 ($\pm$0.004) rad\\
    \hline
    \end{tabular}
    \label{tab:my_label}
\end{table*}

\subsection{Spectroscopy}
\begin{figure*}
\centering
\includegraphics[width=7.8 in]{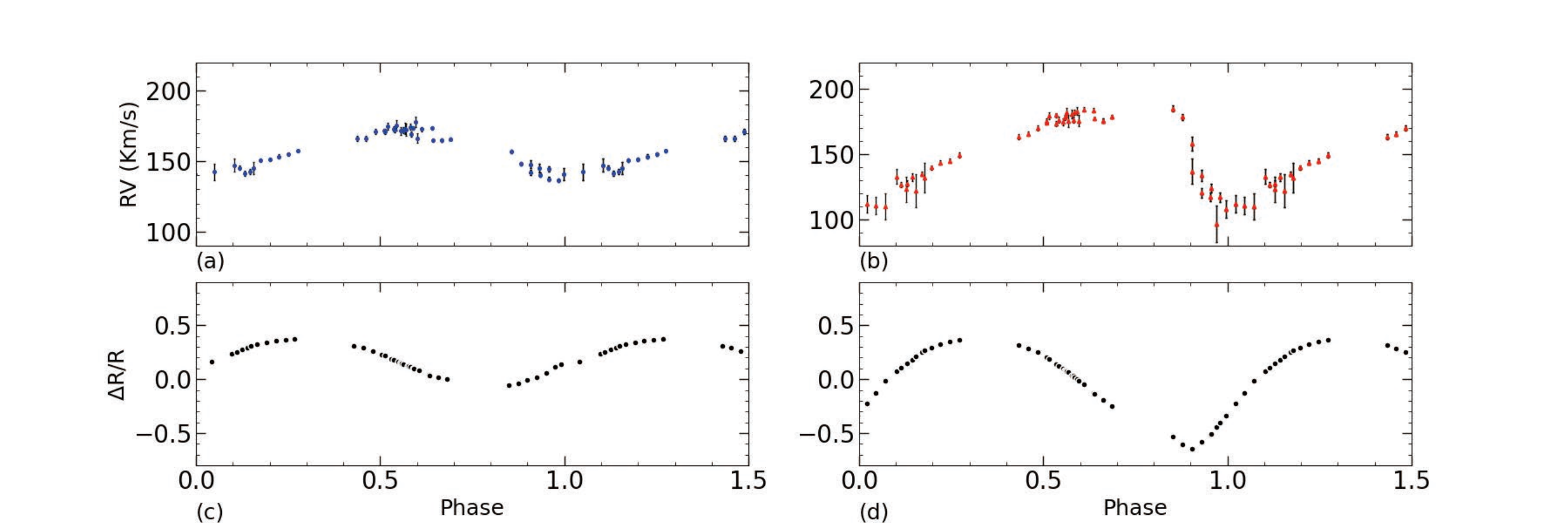}
\caption{The radial velocity curves and their corresponding radius variations. (a)-(b) the radial velocities determined from the blue- and red-arm MRS of LAMOST DR9 are marked in blue and red, respectively. (c)-(d) the radius variations calculated based on the radial velocities estimated from the blue- and red-arm MRS using the Eq~(\ref{eq:5}), respectively.}
\label{fig:RV_Radius}
\end{figure*}

We obtain 50 MRS for EPIC 248846335 from LAMOST DR10, each with a signal-to-noise ratio (S/N) greater than 3.0 in the $i$ band. Those spectra are bifurcated into two wavelength ranges: the red arm covers 630–680 nm, and the blue arm spans 495–535 nm. We adopt the SLAM pipeline \citep{2020RAA....20...51Z} to extract radial velocities (RVs) from the spectra of both arms. However, RV measurements may exhibit systematic discrepancies across different spectrographs and observation nights, potentially reaching several km s$^{-1}$, as documented by \cite{2019RAA....19...75L} and \cite{2020ApJS..251...15Z}. These offsets can be eliminated via the comparison to constant stars \citep{2019RAA....19...75L,2020ApJS..251...15Z}, a technique integrated into the SLAM pipeline. Following this correction method, the computed RVs from the blue and red arms are listed in Tables~\ref{tab:blue_RV} and~\ref{tab:red} and shown in Figure~\ref{fig:RV_Radius}, with panels (a) and (b) illustrating the blue-arm and red-arm RVs, respectively. We base the phase-folding and analysis of the RV curves on the more precise fundamental period derived from the light curve observed by $K2$. The pulsation parameters for the RV curves of both spectral arms are calculated using Eq~(\ref{eq:1}) and listed in the third and fourth columns of Table~\ref{tab:my_label}. The standard deviations of the residuals from the Fourier fits to the RV curves are 5.92 km s$^{-1}$ for the red arm and 2.43 km s$^{-1}$ for the blue arm. \cite{2021ApJS..256...14Z} pointed out that this discrepancy may be attributed to the different precision levels inherent in the red- and blue-arm MRS from LAMOST. Additionally, velocity curves derived from distinct spectral lines, which may reflect disparate kinematics even at identical phases, can account for variations in curve shapes and amplitudes, as suggested by \cite{2021ApJ...919...85B}.

We have collected 92 single-exposure LRS with a signal-to-noise ratio (S/N) exceeding 10.0 from LAMOST DR9 \citep{2021RAA....21..249B}. The MRS survey by LAMOST aims to compile time-series spectra at medium resolution, with the acquisition of radial velocities (RVs) for designated stars being a principal scientific objective \citep{2020ApJS..251...15Z, 2020arXiv200507210L}. The LAMOST LRS survey seeks to determine stellar parameters for a diverse array of targets across the northern hemisphere, specifically those with declinations above -10$^{\circ}$ \citep{2012RAA....12.1243L,2015RAA....15.1095L}; however, it excludes time-domain observations. An investigation by \cite{2019RAA....19...75L} that used multiple MRS observations for nearly 1900 targets revealed that the RV scatter for stars with a standard deviation below 0.5 km s$^{-1}$ was significantly lower—by a factor of 3 to 5—compared to measurements obtained from LRS \citep{2015RAA....15.1095L}. In this paper, we utilize LRS data from LAMOST to analyze the atmospheric parameters of stars. Nonetheless, the determination of these parameters for RRLs from spectral data is contentious. Studies have shown that both low and high-resolution spectra can yield accurate stellar parameters at various phases, and the derived [Fe/H] abundances appear to be phase-independent \citep{2011ApJS..197...29F,2021ApJ...908...20C}. However, \cite{2021ApJ...908...20C} noted that the large amplitude variations in RRLs can systematically alter the effective temperature and luminosity, potentially affecting the determination of chemical abundances if spectra are taken at different phases.

It has been suggested by \cite{2010A&A...519A..64K} that the most favorable phase for spectral analysis corresponds to the maximum radius of RRLs, during which stellar parameters can be precisely determined using the equivalent width method, as implemented in the literature \citep{2014MNRAS.445.4094F,2021MNRAS.506.6117W}. The radius changes of a pulsating star can be inferred from the periodic RV variations using the following equations:
\begin{equation}
    \dot R = -p(V_r(t)-V_*)
\label{eq:4}
\end{equation}
\begin{equation}
    \Delta R(t) = \int_{0}^{P} \dot R dt
\label{eq:5}
\end{equation}
where $P$ is the period, and $V_*$ represents the center-of-mass RV of the star, for which we adopt the mean values of the RV curves in this study. The factor $p$ accounts for the geometrical projection and limb-darkening corrections. We used a value of $p = 1.25$, consistent with that adopted by \cite{2021MNRAS.506.6117W} and based on the investigation of \cite{2017A&A...604A.120N}. The maximum radius variation derived from the RVs of the red-arm MRS of LAMOST, $\Delta$R(t) = 0.37$\pm$0.02 R$_\odot$ observed at phase $\phi_{max}$ = 0.323 $\pm$ 0.003 and that from the RVs of the blue-arm MRS is $\Delta$R(t) = 0.52$\pm$ 0.01 R$_\odot$ observed at phase $\phi_{max}$ = 0.320$\pm$0.004. The maximum radius variations occur at the same phase, within the uncertainties, for both the blue- and red-arm spectra. The atmospheric parameters of the star determined at phase $\phi_{max}$ = 0.319 corresponding to the maximum radius are $T_\mathrm{eff}$ = 6933$\pm$70 K, [Fe/H] = -1.18 $\pm$ 0.14 and logg = 3.35 $\pm$ 0.50 using the template matching method provide by Wang et al (ApJS, 2024, under revision). The radius variations $\Delta R(t)$, derived from RVs and calculated by Eq (4) and (5), are displayed in panels (c) and (d) of Figure~\ref{fig:RV_Radius}.

\begin{longtable*}{cccccc}
\caption{The RVs of the target star measured from the blue-arm MRS of LAMOST DR10. The data is sorted based on the BJD (Barycentric Julian Date) time.}\\\toprule
\label{tab:blue_RV}
ID	&	BJD	&	Phase	&	RV	&	$\sigma$	&	S/N	\\
	&	(day)	&	(rad)	&	(Km/s)	&	(Km/s)	&		\\
(1)	&	(2)	    &	(3)	    &	(4)	   &	(5)	    &	(6)	\\
\hline
\midrule
\endfirsthead
\midrule
ID	&	BJD	&	Phase	&	RV	&	$\sigma$	&	S/N	\\
	&	(day)	&	(rad)	&	(Km/s)	&	(Km/s)	&		\\
(1)	&	(2)	    &	(3)	    &	(4)	   &	(5)	    &	(6)	\\
\hline
\endhead
\midrule
\endfoot
\endlastfoot

1	&	2458183.122	&	0.333	&	140.6	&	4.3	&	3.56	\\
2	&	2458183.155	&	0.384	&	142.3	&	5.7	&	4.38	\\
3	&	2458183.190	&	0.439	&	147.2	&	4.4	&	3.57	\\
4	&	2458183.224	&	0.491	&	145.0	&	4.2	&	3.01	\\
5	&	2458823.396	&	0.901	&	173.2	&	3.9	&	4.27	\\
6	&	2458823.407	&	0.919	&	169.4	&	1.8	&	4.29	\\
7	&	2458823.417	&	0.934	&	166.4	&	3.4	&	3.01	\\
8	&	2458824.389	&	0.452	&	145.2	&	1.6	&	7.12	\\
9	&	2458824.398	&	0.467	&	141.5	&	1.8	&	7.42	\\
10	&	2458824.407	&	0.482	&	142.6	&	1.8	&	6.91	\\
11	&	2458829.341	&	0.191	&	157.0	&	1.3	&	9.91	\\
12	&	2458829.357	&	0.217	&	148.0	&	1.2	&	12.4	\\
13	&	2458829.373	&	0.242	&	142.1	&	1.6	&	13.03	\\
14	&	2458829.390	&	0.268	&	140.3	&	1.6	&	12.69	\\
15	&	2458829.406	&	0.293	&	137.3	&	1.7	&	12.86	\\
16	&	2458829.422	&	0.318	&	136.4	&	1.4	&	11.99	\\
17	&	2458857.276	&	0.846	&	171.4	&	1.3	&	8.07	\\
18	&	2458857.293	&	0.872	&	173.1	&	2.0	&	7.46	\\
19	&	2458857.309	&	0.897	&	171.6	&	1.9	&	7.34	\\
20	&	2458857.325	&	0.923	&	173.7	&	1.3	&	8.47	\\
21	&	2458857.341	&	0.948	&	172.9	&	1.6	&	8.81	\\
22	&	2458857.358	&	0.974	&	173.8	&	1.3	&	8.39	\\
23	&	2458910.172	&	0.508	&	150.6	&	1.0	&	15.31	\\
24	&	2458910.189	&	0.534	&	151.2	&	1.0	&	15.42	\\
25	&	2458910.205	&	0.559	&	153.3	&	1.6	&	10.78	\\
26	&	2458910.222	&	0.585	&	155.0	&	1.1	&	11.71	\\
27	&	2458910.238	&	0.610	&	157.2	&	1.1	&	11.12	\\
28	&	2458941.056	&	0.771	&	166.1	&	2.0	&	8.34	\\
29	&	2458941.072	&	0.796	&	166.3	&	1.9	&	9.47	\\
30	&	2458941.089	&	0.822	&	171.1	&	2.0	&	9.32	\\
31	&	2458941.105	&	0.847	&	170.9	&	1.8	&	10.1	\\
32	&	2458941.121	&	0.873	&	172.1	&	1.7	&	9.23	\\
33	&	2458941.137	&	0.898	&	172.5	&	1.8	&	9.22	\\
34	&	2458950.068	&	0.854	&	174.8	&	2.3	&	6.58	\\
35	&	2458950.084	&	0.879	&	175.7	&	3.7	&	5.01	\\
36	&	2458950.101	&	0.905	&	172.2	&	4.2	&	4.54	\\
37	&	2458950.117	&	0.930	&	177.7	&	3.8	&	4.36	\\
38	&	2459182.379	&	0.891	&	171.4	&	2.5	&	5.51	\\
39	&	2459182.395	&	0.916	&	174.4	&	2.0	&	6.68	\\
40	&	2459238.276	&	0.242	&	147.8	&	2.9	&	7.48	\\
41	&	2459238.292	&	0.267	&	145.2	&	3.2	&	9.3	\\
42	&	2459238.308	&	0.292	&	144.4	&	2.0	&	10.08	\\
43	&	2459634.210	&	0.977	&	164.9	&	1.0	&	15.66	\\
44	&	2459634.226	&	0.000	&	164.8	&	1.1	&	14.64	\\
45	&	2459634.241	&	0.024	&	165.6	&	1.0	&	15.32	\\
\bottomrule
\end{longtable*}

\begin{longtable*}{cccccc}
\caption{The RVs of the target star measured from the red-arm MRS of LAMOST DR10. The data is sorted based on the BJD time.}\\\toprule
\label{tab:red}  
ID	&	HJD	&	Phase	&	RV	&	$\sigma$	&	S/N	\\
	&	(day)	&	(rad)	&	(Km/s)	&	(Km/s)	&		\\
(1)	&	(2)	&	(3)	&	(4)	&	(5)	&	(6)	\\
\hline
\midrule
\endfirsthead
\midrule
ID	&	HJD	&	Phase	&	RV	&	$\sigma$	&	S/N	\\
	&	(day)	&	(rad)	&	(Km/s)	&	(Km/s)	&		\\
(1)	&	(2)	&	(3)	&	(4)	&	(5)	&	(6)	\\
\hline
\endhead
\midrule
\endfoot
\endlastfoot

1	&	2458183.106	&	0.308	&	96.8	&	13.9	&	7.07 	\\
2	&	2458183.122	&	0.333	&	108.1	&	6.7	&	7.06 	\\
3	&	2458183.139	&	0.359	&	112.1	&	6.4	&	7.70 	\\
4	&	2458183.155	&	0.384	&	110.7	&	6.6	&	8.23 	\\
5	&	2458183.171	&	0.409	&	110.0	&	10.2	&	4.64 	\\
6	&	2458183.190	&	0.439	&	132.9	&	5.6	&	5.90 	\\
7	&	2458183.207	&	0.465	&	123.2	&	9.8	&	4.98 	\\
8	&	2458183.224	&	0.491	&	121.9	&	12.8	&	4.39 	\\
9	&	2458183.240	&	0.516	&	132.1	&	11.4	&	3.71 	\\
10	&	2458823.396	&	0.901	&	182.2	&	3.3	&	8.61 	\\
11	&	2458823.407	&	0.919	&	176.3	&	2.3	&	8.67 	\\
12	&	2458823.417	&	0.934	&	175.7	&	4.3	&	6.74 	\\
13	&	2458824.389	&	0.452	&	126.5	&	2.2	&	10.89 	\\
14	&	2458824.398	&	0.467	&	127.4	&	2.3	&	11.88 	\\
15	&	2458824.407	&	0.482	&	132.5	&	3.1	&	11.20 	\\
16	&	2458829.341	&	0.191	&	184.9	&	2.2	&	16.59 	\\
17	&	2458829.357	&	0.217	&	178.4	&	2.2	&	19.62 	\\
18	&	2458829.373	&	0.242	&	158.2	&	5.2	&	20.25 	\\
19	&	2458829.390	&	0.268	&	133.7	&	4.2	&	19.27 	\\
20	&	2458829.406	&	0.293	&	124.1	&	3.2	&	19.69 	\\
21	&	2458829.422	&	0.318	&	117.0	&	3.5	&	15.70 	\\
22	&	2458857.276	&	0.846	&	174.2	&	1.7	&	14.23 	\\
23	&	2458857.293	&	0.872	&	173.2	&	1.8	&	12.88 	\\
24	&	2458857.309	&	0.897	&	177.8	&	1.9	&	12.63 	\\
25	&	2458857.325	&	0.923	&	182.2	&	1.9	&	15.00 	\\
26	&	2458857.341	&	0.948	&	184.3	&	1.5	&	14.94 	\\
27	&	2458857.358	&	0.974	&	183.5	&	1.6	&	14.12 	\\
28	&	2458910.172	&	0.508	&	134.8	&	1.6	&	22.30 	\\
29	&	2458910.189	&	0.534	&	139.7	&	1.7	&	22.20 	\\
30	&	2458910.205	&	0.559	&	143.4	&	1.7	&	16.40 	\\
31	&	2458910.222	&	0.585	&	144.8	&	1.6	&	18.21 	\\
32	&	2458910.238	&	0.610	&	149.3	&	1.8	&	17.19 	\\
33	&	2458941.056	&	0.771	&	163.4	&	1.9	&	11.81 	\\
34	&	2458941.072	&	0.796	&	165.5	&	1.7	&	14.19 	\\
35	&	2458941.089	&	0.822	&	170.0	&	1.8	&	14.65 	\\
36	&	2458941.105	&	0.847	&	175.2	&	1.9	&	15.81 	\\
37	&	2458941.121	&	0.873	&	179.7	&	1.8	&	14.56 	\\
38	&	2458941.137	&	0.898	&	179.1	&	1.8	&	15.12 	\\
39	&	2458950.068	&	0.854	&	179.4	&	2.6	&	9.96 	\\
40	&	2458950.084	&	0.879	&	176.1	&	2.4	&	7.59 	\\
41	&	2458950.101	&	0.905	&	175.5	&	4.6	&	6.70 	\\
42	&	2458950.117	&	0.930	&	182.9	&	3.2	&	6.32 	\\
43	&	2459182.379	&	0.891	&	174.5	&	2.8	&	8.79 	\\
44	&	2459182.395	&	0.916	&	180.5	&	3.4	&	10.62 	\\
45	&	2459238.276	&	0.242	&	136.9	&	9.7	&	12.19 	\\
46	&	2459238.292	&	0.267	&	120.4	&	3.7	&	15.00 	\\
47	&	2459238.308	&	0.292	&	117.5	&	3.5	&	16.33 	\\
48	&	2459634.210	&	0.977	&	177.3	&	1.5	&	19.14 	\\
49	&	2459634.226	&	0.000	&	175.7	&	1.5	&	17.21 	\\
50	&	2459634.241	&	0.024	&	179.0	&	1.3	&	18.26 	\\

\bottomrule
\end{longtable*}

\section{Numerical Modeling}
\subsection{Parameter Calculation}
To calculate the bolometric luminosity of the star, the calibrated distance of the star provided by Gaia DR3 as d = 6839.27$_{-789.22}^{+1327.24}$ pc is used. The formula given by \cite{2018AJ....156...58B} is adopted as follows,
\begin{equation}
\centering
M = M_G + 5(1-\log d) - A_G
\label{eq:6}
\end{equation}

\begin{equation}
-2.5\log L = M + BC_G (T_{\rm eff}) - M_{bol,\odot}
\label{eq:7}
\end{equation}
where $M$ and $M_{G}$ = 14.7 mag are the absolute magnitude and apparent magnitude in $G$ band of Gaia, respectively. The extinction coefficient A$_{G}$ in formula~(\ref{eq:5}) is 0.1224 in $G$ band \citep{2021AJ....161..147B}. The parameter M$_{bol,\odot}$ in formula~(\ref{eq:6}) is the bolometric magnitude of the Sun, which is defined by IAU and the value is 4.74 mag \citep{2015arXiv151006262M}. BC$_G$($T_{\rm eff}$) is the bolometric correction which depends only on the effective temperature \citep{2018A&A...616A...8A}. The bolometric luminosity estimated for this star is $L$ = 49.70$_{-1.80}^{+2.99}$ L$_\odot$. The metallicity of the star is calculated by adopting the following equations,
\citep{2012MNRAS.427..127B},
\begin{equation}
\centering
[Fe/H] = \log (Z/X)-\log (Z/X)_\odot 
\label{eq:8}
\end{equation}

\begin{equation}
\centering
Y = 0.2485+1.78 Z
\label{eq:9}
\end{equation}

\begin{equation}
\centering
X + Y + Z = 1
\label{eq:10}
\end{equation}
where the value of (Z/X)$_\odot$ is 0.0207 \citep{2011SoPh..268..255C}. X, Y, and Z are the hydrogen, helium, and metal abundance by the mass fraction of the star, which we estimate as X = 0.748 $\pm$ 0.001, Y = 0.250 $\pm$ 0.001 and Z = 0.0010 $\pm$ 0.0003, respectively. The value of (Z/X)$_\odot$ in RSP inlist provided by \cite{2009ARA&A..47..481A} is different from that value of \cite{2011SoPh..268..255C}. The mass of the star is calculated as M = 0.56 $\pm$ 0.07 $M_\odot$ using the Eq (22) of \cite{1998A&A...333..571J}, which is based on the horizontal branch models that indicate the dependence of the stellar mass on the metallicity within the instability strip proposed by \cite{1991ApJS...76..911C}. In this work, we only use the mass as the initial mass to construct the grid of models.

\subsection{Model construction and Selection}
The stellar radial pulsation convective code based on the time-dependent turbulent convection model \citep{1986A&A...160..116K} was implemented by \cite{2008AcA....58..193S}. This model can effectively reproduce the light curves and RV curves of classical pulsating variables as it combines the convection and the pulsation driven by partial ionization. The turbulent energy and the kinetic energy are coupled to each other through coupling terms \citep{2008AcA....58..193S}, which are controlled by the eight order of the unity convection parameters as the mixing-length $\alpha$, the eddy-viscous dissipation $\alpha_m$, the turbulent source $\alpha_s$, the convective flux $\alpha_c$, the turbulent dissipation $\alpha_d$, the turbulent pressure $\alpha_p$, the turbulent flux $\alpha_t$ and the radiative cooling $\gamma_r$. According to \cite{2019ApJS..243...10P}, the convection parameters for modeling different types of stars, for instance, Cepheids, RRLs, and other stellar systems, slightly different values should be considered in constructing models. They suggested that $\alpha_t$ $\simeq$ 0.01, $\alpha_m$ $\lesssim$ 1.0, and $\alpha$ $\lesssim$ 2 are useful initial choices in experience. The investigation of \cite{2023MNRAS.521.4878K} revealed that varying convective parameters have distinct effects on the final radial velocity and light curves as presented in their Figure 2 and 3. They pointed out that among parameters $\alpha_m$ of RSP has the most significant effect on the resulting radial velocity and light curves, while other parameters have little effect. We adjust the values of $\alpha_m$ and the other parameters following those recommended by \cite{2019ApJS..243...10P}. The value sets of these parameters are listed in Table ~\ref{tab:convince} to produce optimal RV and light curves of the star, and the values of convective parameters are fixed to 4 sets given in Table~\ref{tab:convince}.

\begin{table}
    \centering
    \renewcommand\tabcolsep{12pt}
    \caption{The convection parameter sets A, B, C and D.}
    \begin{tabular}{lcccc}
    \hline
    \hline
   Parameter & Set A & Set B & Set C & Set D\\
    \hline
       $\alpha$&1.5  &1.5  &1.5 &1.5\\
       $\alpha_m$&0.8&0.65 &0.8 &0.6\\
       $\alpha_s$&1.0&1.0 &1.0 &1.0\\
       $\alpha_c$&1.0&1.0 &1.0 &1.0\\
       $\alpha_d$&1.0&1.0 &1.0 &1.0\\
       $\alpha_p$&0.0&0.0 &1.0 &1.0\\
       $\alpha_t$&0.00&0.00 &0.01 &0.01\\
       $\gamma_r$&0.0&1.0 &0.0 &1.0\\
    \hline
    \end{tabular}
    \label{tab:convince}
\end{table}

In this study, a grid of models is calculated using the RSP module of MESA with the stellar parameters. As suggested by \citet{2019ApJS..243...10P}, the initial input parameters mass, luminosity, effective temperature, hydrogen abundance (X), and metal abundance (Z) can be freely chosen and do not necessarily need to originate from a MESAstar model. Based on the atmospheric parameters determined from the LRS of LAMOST, the effective temperature of the star varies within the range of 6100--7300 K, which falls within the typical effective temperature range of RRLs. The metallicity determined from those spectra ranges from -1.72 to -0.53. However, the observed phases of the LRS do not cover the entire pulsation cycle. Therefore, the set of metallicity used in this work is adjusted to -2.90 $\leq$ [Fe/H] $\leq$ 0.0. The metal abundance Z and hydrogen abundance X are calculated using Eq (8), (9), and (10), based on the corresponding values of [Fe/H]. This method was also adopted by \citet{2021MNRAS.506.6117W} for determining the sets of Z and X in constructing their model for the non-Blazhko RRab star EZ Cnc (EPIC 212182292). The absolute luminosity and mass determined in this work are $L$ = $49.76_{-1.80}^{+2.99}$ L$_\odot$ and M = 0.56 $\pm$ 0.02 M$_\odot$, respectively. To derive the optimal model of this star, we consider a wide range of luminosity and mass sets: 40 $\leq$ $L$/L$_\odot$ $\leq$ 65 and 0.35 $\leq$ M/M$_\odot$ $\leq$ 0.75. The resolution of the grid is set as $\Delta_{M/M_\odot}$ = 0.01, $\Delta_{T_\mathrm{eff}}$ = 50 K, $\Delta_{L/L_\odot}$ = 1, and $\Delta_{\mathrm{[Fe/H]}}$ = 0.1. An exemplary list is included in the appendix of our paper. The absolute value $\Gamma$ = 4.13 $\times$ 10$^{-6}$ is adopted to ensure that the models converge to a full amplitude solution, as documented by \citet{2019ApJS..243...10P}.

In our analysis, we compare the nonlinear periods derived from the models with the fundamental period determined from the observed light curve. An uncertainty of $\Delta P$ = 0.0007 days is applied for the fundamental period to ensure that the differences between the main period obtained from the $K2$ light curve and the periods derived from the models are smaller than this uncertainty value. This criterion helps us select the appropriate models from the grid. We obtain 40 models that meet the period criterion.

\begin{figure*}
	\centering
\includegraphics[width=0.9\textwidth]{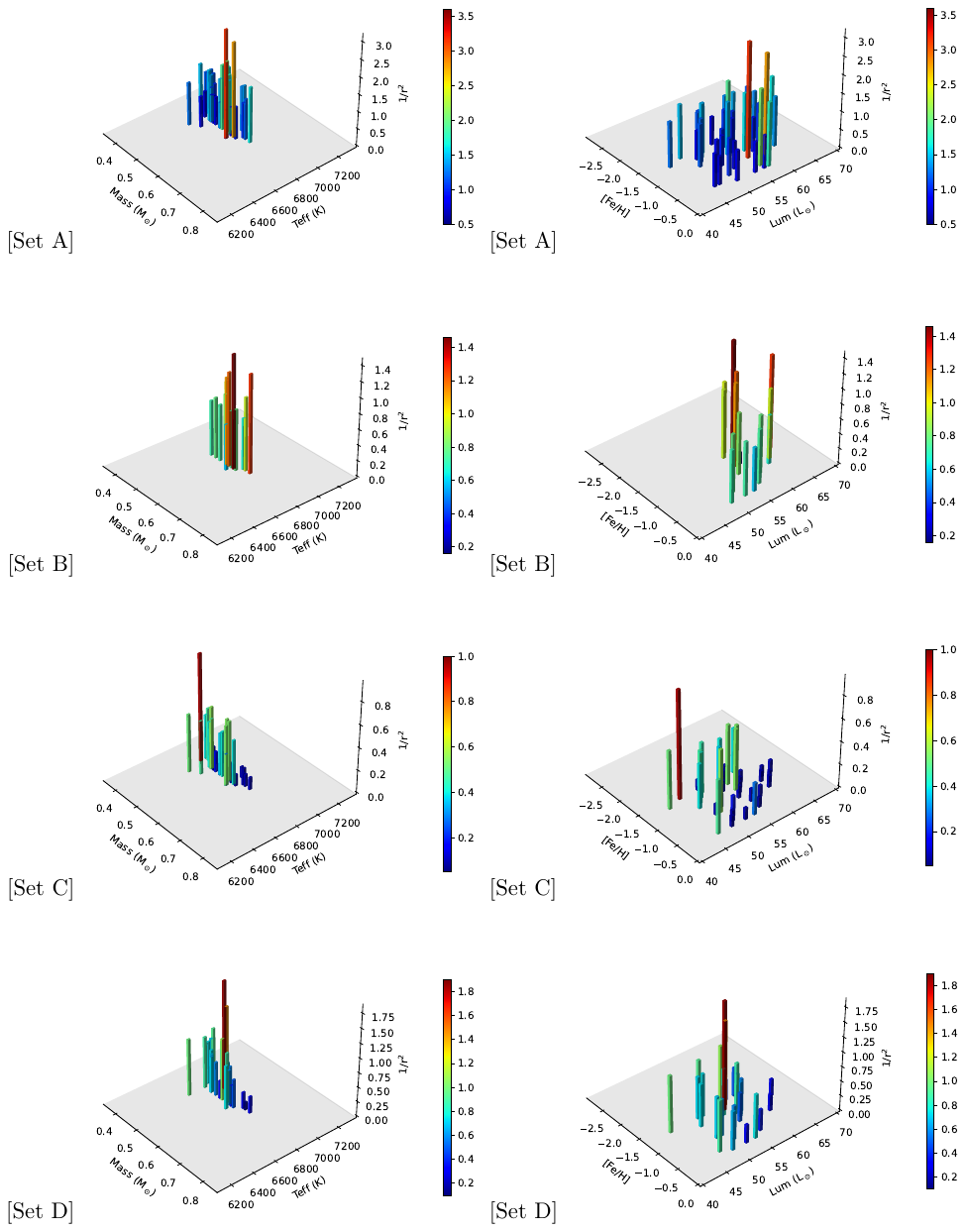}	

	\caption{Determination of the best fit model (residual $r$) of SetA, SetB, SetC and SetD. For better visibility, Panel (a),(c),(e) and (g): $1/r^2$ are plotted as a function of mass (m$_\odot$) and $T_\mathrm{eff}$ with the corresponding color scale for the different four convective parameter sets, respectively. Panel (b),(d),(f), and (h): $1/r^2$ are plotted as a function of metallicity ([Fe/H]) and luminosity (L$_\odot$)  with the corresponding color scale for the different four convective parameter sets, respectively.}
	\label{fig:SetA_d}
\end{figure*}

The light curves of the 40 models are generated using four different convection parameter sets. The convection parameters are adjusted to generate optimal light curves and radial velocity (RV) curves of the star. To maintain consistency between the model light curves and the $K2$ light curve, we convert the model light curves to the $Kepler$ white band using the bolometric calibration coefficient \citep{2019MNRAS.489.1072L}, which depends only on the effective temperature. The residuals $r$ between the models and the observations in the Fourier parameter space \citep{2013MNRAS.428.3034S} are calculated using the following equation:

\begin{equation}
r = \sqrt{\sum{\frac{(p_{i,\mathrm{mod}}-p_{i,\mathrm{obs}})^2}{p_{i,\mathrm{obs}}^2}}}
\end{equation}

where $p_i$ represents one of the low-order Fourier parameters and amplitudes, $p$ $\in$ $\lbrace$RT, R$_{21}$, R$_{31}$, $\phi_{21}$, $\phi_{31}$ $\rbrace$, $p_{i,\mathrm{mod}}$ refers to our models, and $p_{i,\mathrm{obs}}$ refers to the observed curves. The smaller the value of $r$, the closer the observed $K2$ light curves and LAMOST RV curves are to those modeled with RSP and MESA, respectively. The distribution of $r$ in the space of the stellar parameters for SetA, SetB, SetC, and SetD is presented in Figure~\ref{fig:SetA_d}.

It should be noted that not all 40 models converge for each set of convection parameters. We obtain one model from each convection parameter set with the smallest residual $r$ value. The stellar parameters of the four models are listed in Table~\ref{tab:model}. The $\sigma_{\mathrm{mod,RVC}}$ and $\sigma_{\mathrm{mod,LC}}$ are the standard errors of the residuals between the observed RV curves and light curves and their corresponding model-derived curves, respectively. Only the model derived from the convection parameter SetA satisfies $\frac{\sigma_{\mathrm{mod,RVC}}}{\sigma_{\mathrm{obs,RVC}}}$ $\leq$ 3 and $\frac{\sigma_{\mathrm{mod,LC}}}{\sigma_{\mathrm{obs,LC}}}$ $\leq$ 3, suggesting that it is the optimal model of the star.

We estimate the uncertainties of the best-fitting model parameters using the prescription derived by \citet{1986ApJ...305..740Z}, a method commonly adopted in the literature \citep{2008MNRAS.385..430C,2012MNRAS.420.1462R,2019MNRAS.486.3560F}. The equation is as follows:

\begin{equation}
\sigma = d^2/(S-S_0),
\end{equation}

where $\sigma$ is the uncertainty of the parameter, $d$ is the step size of the parameter within the model grid, $S_0$ is the $r^2$ value of the best-fitting model (i.e., the minimum value), and $S$ is the $r^2$ value for the model with the prescribed change of the parameter by the amount $r$ while keeping all other parameters fixed. The best-fitting model parameters and their uncertainties are: $M$ = 0.59 $\pm$ 0.05 M$_\odot$, $T_\mathrm{eff}$ = 6700 $\pm$ 220 K, [Fe/H] = -1.2 $\pm$ 0.2, and $L$ = 56.0 $\pm$ 4.2 L$_\odot$. The projection factor values of the star are determined to be 1.20 $\pm$ 0.02 and 1.59 $\pm$ 0.13, which are constrained by the blue- and red-arm observed velocities and their corresponding RV curves derived from the structural profiles of the optimal model.

We also calculate the light and RV curves of the optimal model considering different mesh numbers (e.g., RSP\_nz=150, RSP\_nz\_outer=30, and RSP\_nz=200, RSP\_nz\_outer=60) and time steps per pulsation cycle (RSP\_target\_steps\_per\_cycle=200 and RSP\_target\_steps\_per\_cycle=600). The results indicate that the light and RV curves of the models are not sensitive to these parameters, consistent with the findings of \citet{2019ApJS..243...10P}.
\begin{table*}
\renewcommand\tabcolsep{6.0pt}
\centering
\caption{Properties of the best models for EPIC 248846335 in the four convection parameters. The model number (Column 1), the stellar parameters of different models (Columns 2-7), the surface gravity (Column 8), the offset values $d$ of different models (Column 9), the standard errors of the light curve and radial velocity curve (Columns 10-12), the different convective parameter sets (Column 13). }
\begin{tabular}{lcccccccccccc}
\hline
\hline
Model	&	mass	&	Lum	&	$T_\mathrm{eff}$	&	X	&	Z	&	[Fe/H]	& log $g$	&r	&	$\sigma_{LC}$	&	$\sigma_{red,rv}$	&	$\sigma_{blue,rv}$	&	Set	\\
	&	(m$_\odot$)	&	(L$_\odot$)	&	(K)	&		&		&	&	&		&	(mag)	&	(Km/s)	&	(Km/s)	&		\\
(1)	&	(2)	&	(3)	&(4)		&(5)		&(6)		&(7)	& (8)	&	(9)	&(10)		&(11)		&(12)		&	(13)	\\
\hline
1	&	0.59	&	56	&	6700	&	0.7488	&	0.0010	&	-1.20 	& 2.65	&0.56	&	0.011	&	5.85	&	6.03	&	Set A	\\
2	&	0.60	&	59	&	6750	&	0.7498	&	0.0006	&	-1.40 	& 2.62	&0.83	&	0.027	&	10.20	&	3.92	&	Set B	\\
3	&	0.45	&	47	&	6750	&	0.7498	&	0.0006	&	-1.40 	& 2.65	&1.02	&	0.020	&	10.90	&	4.52	&	Set C	\\
4	&	0.51	&	55	&	6850	&	0.7488	&	0.0010	&	-1.20 	& 2.63	&0.73	&	0.029	&	7.17	&	4.69	&	Set D	\\
\hline
\end{tabular}
\label{tab:model}
\end{table*}

\begin{figure}
    \centering
    \includegraphics[width=3.2 in]{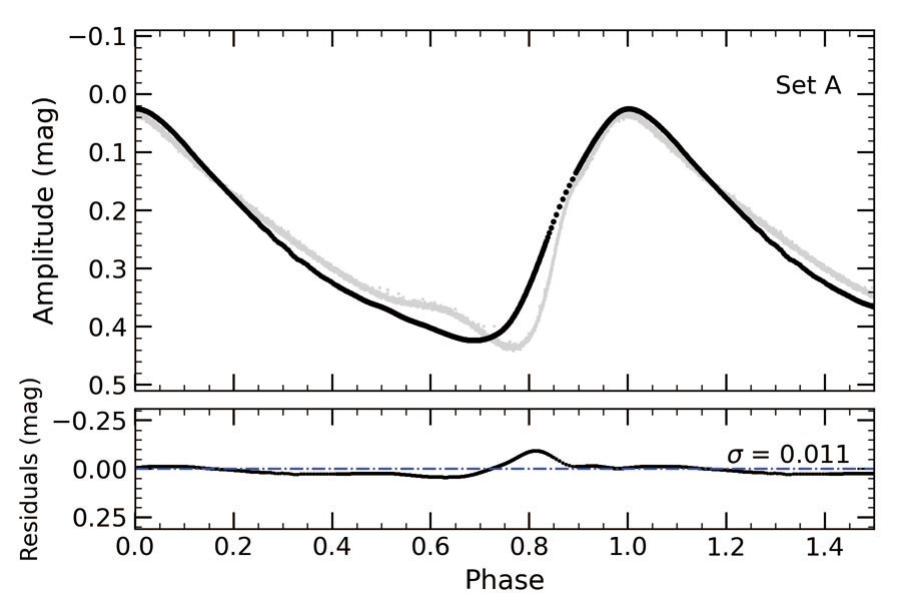}
    \caption{Upper panel: comparison between the light curve (gray) observed by $K2$ and that produced by the RSP modules of MESA. Bottom panel: the residuals of this comparison.}
    \label{fig:SetA_LC}
\end{figure}

\begin{figure}
    \centering
    \includegraphics[width=3.2 in]{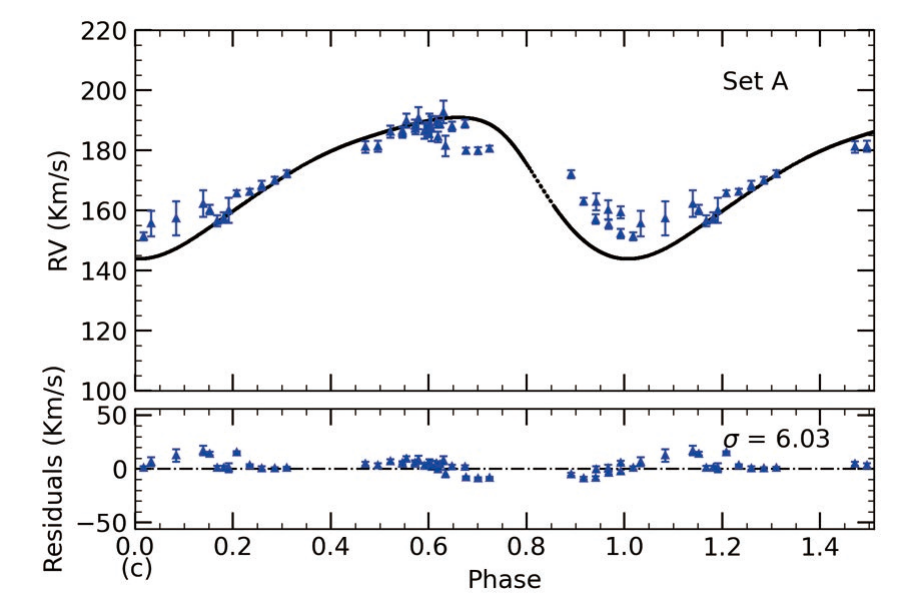}
    \caption{Upper panel: comparison between the RV curves (blue) provided by the blue-arm MRS of LAMOST DR9 and that produced by the RSP module of MESA. Bottom panel: the residuals of this comparison. }
    \label{fig:SetA_RV_Blue}
\end{figure}

\begin{figure}
    \centering
    \includegraphics[width=3.2 in]{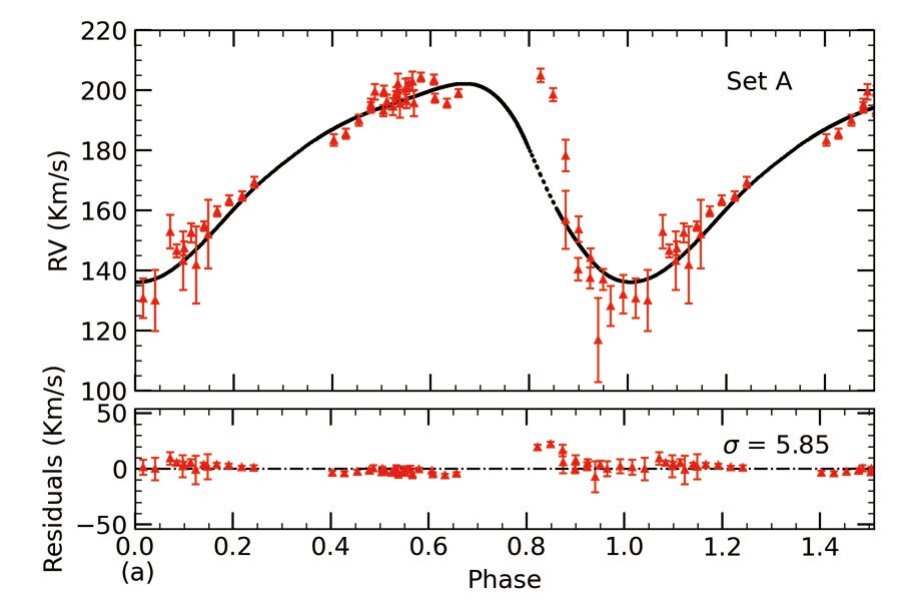}
    \caption{Upper panel: comparison between the RV curves (red) provided by the red-arm MRS of LAMOST DR9 and that produced by the RSP module of MESA. Bottom panel: the residuals of this comparison.}
    \label{fig:SetA_RV_Red}
\end{figure}

\section{Discussion}
\begin{figure}
   \centering
   \includegraphics[width=3.6 in]{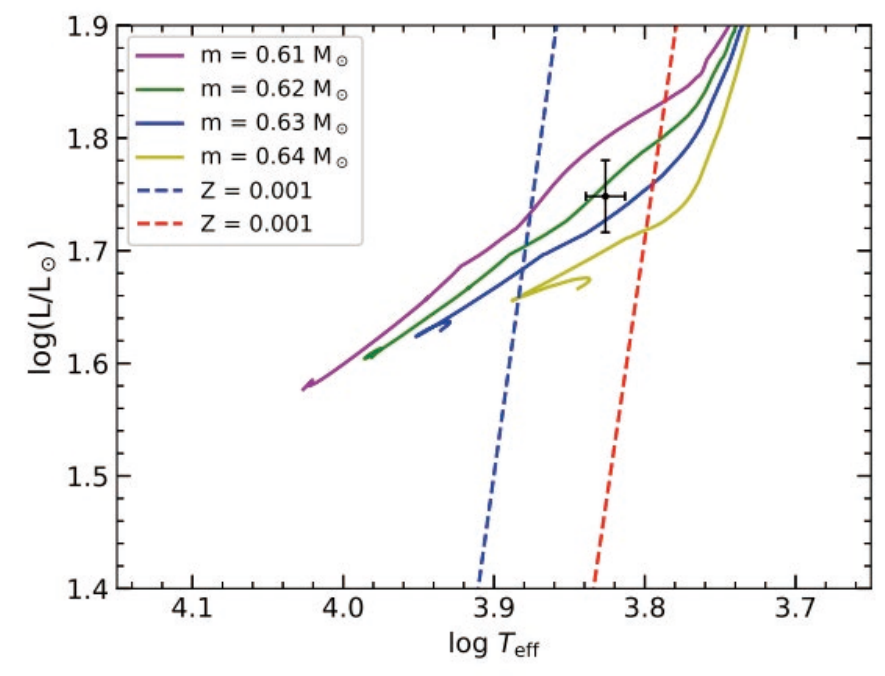}
    \caption{Hertzsprung-Russell diagram for comparison between the best-matching model to the stars EPIC 248846335 and the HB models. Masses of models and evolutionary tracks are color-coded as indicated in the legend. The black pentagon is the optimal model of Set A given in Table~\ref{tab:model}. The blue and red dashed lines represent the blue and red edges of the instability strip \citep{2019AstL...45..353F}, respectively.}
    \label{fig:tracks}
\end{figure}

\citet{2021ASPC..529...63C} pointed out that modeling the evolution of RR Lyrae stars is not a trivial task, which includes difficulties related to uncertainties in modeling the helium flash and mass loss on the red giant branch. We adopt the updated horizontal branch (HB) models from the Bag of Stellar Tracks and Isochrones (BaSTI) project \citep{2018ApJ...856..125H} to calculate the properties of HB models (M = 0.61 $M_\odot$, M = 0.62 $M_\odot$, M = 0.63 $M_\odot$, and M = 0.64 $M_\odot$) using a chemical composition of $Z$ = 0.001 and $Y$ = 0.246, which is similar to that derived from the optimal model, with the input parameter $\alpha$ = 1.5 and the mass loss efficiency $\eta$ = 0.4 \citep{1975MSRSL...8..369R}. The comparison between the pulsation modeling results and the evolutionary tracks is presented in the H-R diagram (Figure~\ref{fig:tracks}), where the optimal model is located in the middle of the instability strip. The comparison between the positions of the optimal model and the evolutionary tracks indicates discrepancies between the masses and luminosities derived from pulsation modeling and those associated with the evolutionary tracks situated in analogous regions of the instability strip. \citet{2011MNRAS.417.1022N}, who studied 19 non-Blazhko RRab stars using $Kepler$ photometry, revealed a discrepancy between the masses and luminosities derived from evolutionary tracks and those obtained from pulsation calculations. As they documented, it is not clear whether the luminosity and mass derived from the evolutionary tracks or those calculated from the pulsation code are correct. In a study conducted by \citet{2021MNRAS.506.6117W}, a comparison between results obtained from evolutionary tracks and pulsation modeling revealed a similar discrepancy in masses and luminosities. \citet{2022MNRAS.515.3439N} suggested that there is no direct mass determination for any known RRLs since no RRL is known to be in an eclipsing binary system. The most promising candidate for a RRL in a binary system turned out to be a star with a significantly smaller mass and formed through a different evolutionary channel \citep{2012Natur.484...75P}. The search for RRLs in binary systems is ongoing and has resulted in several candidates \citep{2018pas6.conf..248H}. Unfortunately, binary systems detected using the light-time effect will not yield dynamical masses for RRLs. However, \citet{2022MNRAS.515.3439N} also suggested that in the absence of direct mass determination for RRLs, we may compare the mass estimates with those determined using various methods, such as those based on the shape of the light curve \citep{1993ApJ...410..526S}, comparison with evolutionary tracks \citep{2019ARep...63..203M}, or asteroseismic modeling \citep{2015MNRAS.452.4283M}.

We determine two different values of the projection factors for the star, possibly attributed to the fact that the observed radial velocity (RV) curves are obtained from different spectral lines of the MRS of LAMOST \citep{2020RAA....20...51Z}. The study by \citet{2012A&A...543A..55N} has shown that differences in radial velocity measurements may change the determination of the projection factor for $\delta$ Cephei. \citet{2017A&A...604A.120N} determined the values of the projection factors ranging from 1.273 to 1.329 using different amplitudes of RV curves estimated from 17 different spectral lines. 
The study of \citet{2019A&A...623A.109G} had also revealed that the RV curves derived from different atmospheric layers of the stars have different amplitudes as they extracted the radial velocities of the star RR Lyrae from the sodium and the H$\alpha$ absorption lines corresponding to the deep layers of the photosphere and upper atmosphere, respectively. As previous mentioned, the radial velocity curves derived from the blue and red arm MRS of LAMOST are based on the Mg Ib triple and H$\alpha$ lines  \citep{2020RAA....20...51Z}, respectively. 

The amplitudes of the RVs derived from the red- and blue-arm MRS are 66.10 ($\pm$5.26) km s$^{-1}$ and 43.53 ($\pm$2.01) km s$^{-1}$, respectively, as listed in Table~\ref{tab:my_label} in our paper. The former amplitude is 51\% larger than the latter one, which is consistent with the results in the literature of \citet{2021ApJ...919...85B}. In their study, the RV amplitudes derived from H$\alpha$ are 24\%-52\% larger than the amplitudes derived from the Mg Ib triplet, as listed in Tables 8 and 9 of their paper. \citet{2020ApJ...896L..15B} suggested that the amplitude difference of RV curves is caused by the physical conditions under which the spectral lines form. According to their findings, a smaller optical depth corresponds to a larger RV amplitude. \citet{2021ApJ...919...85B} suggested that RRLs are pulsating stars, and different lines may exhibit distinct kinematics even when observed at the same phase. As a result, the velocity curves derived from different lines can display varying shapes and amplitudes. 


\section{Conclusions}
In this work, we conduct an astroseismological investigation of the non-Blazhko RRab star EPIC 248846335 using homogeneous Medium-Resolution spectra (MRS) in red and blue arms collected by the LAMOST-$Kepler/K2$ project, along with photometric data provided by the $Kepler$ space telescope. The radial velocity (RV) curves of this star are obtained from the red- and blue-arm spectra of LAMOST DR10. The Fourier decomposition method is applied to the light curve and RV curves to determine the pulsation parameters of the star. The stellar atmospheric parameters, including the effective temperature $T_\mathrm{eff}$ = 6933 $\pm$ 70 K, surface gravity log $g$ = 3.35 $\pm$ 0.50, and metallicity [Fe/H] = -1.18 $\pm$ 0.14, are estimated from the single-exposure Low-Resolution spectra (LRS) of LAMOST DR9. The stellar mass M=0.56 $\pm$ 0.07 $M_\odot$ is also calculated based on the value of [Fe/H]. We determine the absolute luminosity $L$ = $49.70_{-1.80}^{+2.99}$ $L_\odot$ of the star using the distance provided by Gaia DR2.

A series of time-independent convection grid models are constructed based on the estimated stellar parameters using the RSP module of MESA. The fundamental period of the star and the residuals $r$ of the Fourier parameters between the models and observations serve to select the optimal model. The stellar parameters of the optimal model are determined as follows: $T_\mathrm{eff}$ = 6700 $\pm$ 220 K, log $g$ = 2.70, [Fe/H] = -1.20 $\pm$ 0.2, M = 0.59 $\pm$ 0.05 $M_\odot$, and $L$ = 56.0 $\pm$ 4.2 $L_\odot$. The values of the projection factors of the star are constrained to be 1.20 $\pm$ 0.02 and 1.59 $\pm$ 0.13 by the blue- and red-arm observed velocities with their corresponding RV curves derived from the best-fit model. In the future, a larger amount of precise light curves and spectra of RR Lyrae stars (RRLs) will hopefully be obtained, which would bring new constraints to the hydrodynamic models constructed for these stars and help improve our understanding of the stellar evolution of RRLs.


\section*{Acknowledgments}.

We acknowledge the support from the National Natural Science Foundation of China (NSFC) through grants 12090040, 12090042, and 11833002. The Guoshoujing Telescope (the Large Sky Area Multi-object Fiber Spectroscopic Telescope, LAMOST) is a National Major Scientific Project built by the Chinese Academy of Sciences. This work is supported by the International Centre of Supernovae, Yunnan Key Laboratory (No. 202302AN360001). The authors gratefully acknowledge the Kepler team and all who have contributed to making this mission possible. We acknowledge Dr. Tian-Qi Cang and Dr. Jiangtao Wang for providing valuable suggestions regarding the estimation of the atmospheric parameters from the LRS and discussion in this paper. The authors gratefully acknowledge the referee, who gave us very useful suggestions for this paper.

\software{astropy \citep{2013A&A...558A..33A,2018AJ....156..123A,2022ApJ...935..167A}, LightKurve \citep{2021zndo...1181928B, 2021ApJ...922....2D}, Period04 \citep{2005CoAst.146...53L}}

%





\appendix
$\&$star\_job

show$\_$log$\_$description$\_$at$\_$start = .false.

create\_RSP\_model = .true.

save\_model\_when\_terminate = .true.
      
save\_model\_filename = 'final.mod'

initial\_zfracs = 6

color\_num\_files=2
      
color\_file\_names(2)='blackbody\_johnson.dat'
      
color\_num\_colors(2)=5

set\_initial\_age = .true.
      
initial\_age = 0

set\_initial\_model\_number = .true.
      
initial\_model\_number = 0
      
set\_initial\_cumulative\_energy\_error = .true.
      
new\_cumulative\_energy\_error = 0d0
   
$/$ $!$ end of star\_job namelist

$\&$eos
$/$ $!$ end of eos namelist

$\&$kap
   
Zbase = 0.0014d0

kap\_file\_prefix = 'a09'
    
kap\_lowT\_prefix = 'lowT\_fa05\_a09p'
    
kap\_CO\_prefix = 'a09\_co'

! end of kap namelist

$\&$controls

! limit max\_model\_number as part of test\_suite
   
!max\_model\_number = 1000000

$!$ RSP controls

! x\_integer\_ctrl(1) = 10 ! which period to check
   
x\_ctrl(1) = 0.639906d0 ! expected period (in days)

RSP\_mass = 0.65d0
   
RSP\_Teff = 6700d0
   
RSP\_L = 45d0
   
RSP\_X = 0.75d0
   
RSP\_Z = 0.0014d0
   
$!$ parameters for equations
   
RSP\_alfa  = 1.5d0  ! mixing length; alfa = 0: purely radiative model.
   
RSP\_alfam = 0.85d0 ! eddy viscosity; Chi $\&$ Eq $\sim$ RSP\_alfam
   
RSP\_alfas  = 1.0d0
   
RSP\_alfac  = 1.0d0
   
RSP\_alfad  = 1.0d0
   
RSP\_alfap  = 1.0d0
   
RSP\_alfat  = 0.01d0
   
RSP\_gammar  = 1.0d0

RSP\_target\_steps\_per\_cycle = 200
   
RSP\_kick\_vsurf\_km\_per\_sec = 4.5d0 ! can be negative
   
RSP\_fraction\_1st\_overtone = 0d0
   
RSP\_fraction\_2nd\_overtone = 0d0
   
RSP\_nz = 150 ! total number of zones in static model
   
RSP\_nz\_outer = 30 ! number of zones in outer region of static model
   
RSP\_T\_anchor = 11d3 ! approx temperature at base of outer region
   
RSP\_max\_num\_periods = 3000
   
!RSP\_T\_inner = 2d6
  
! output controls
   
terminal\_show\_age\_units = 'days'

!num\_trace\_history\_values = 2
      
trace\_history\_value\_name(1) = 'rel\_E\_err'
      
trace\_history\_value\_name(2) = 'log\_rel\_run\_E\_err'

photo\_interval = 1000
      
profile\_interval = 1
      
history\_interval = 1
      
terminal\_interval = 4000

/ ! end of controls namelist

\&pgstar

!pause = .true.

pgstar\_interval = 6
         
Grid2\_win\_flag = .true.

Grid2\_title = $'4.165 M(2281) \Gamma Z=0.007Classical Cepheid'$
      
History\_Panels1\_xaxis\_name = 'star\_age\_day'
     
History\_Panels\_max\_width = 365 ! only used if $>$ 0.  causes xmin to move with xmax.

!  Grid2\_file\_flag = .true.
      
file\_digits = 7
      
Grid2\_file\_dir = 'png'
      
Grid2\_file\_prefix = 'grid'
      
Grid2\_file\_interval = 5 ! output when mod(model\_number,Grid2\_file\_interval)==0
      
!Profile\_Panels1\_show\_grid = .true.

Profile\_Panels1\_xaxis\_name = 'logtau'
      
Profile\_Panels1\_xaxis\_reversed = .true.
      
Profile\_Panels1\_xmin = -101D0
      
Profile\_Panels1\_xmax = -101D0

Profile\_Panels1\_dymin(4) = 0.02
      
Profile\_Panels1\_yaxis\_name(2) = 'avg\_charge\_He' 

! end of pgstar namelist


\bibliography{sample631}{}
\bibliographystyle{aasjournal}



\end{document}